\documentclass[pdflatex,sn-mathphys-num,iicol]{sn-jnl}
\usepackage{graphicx}%
\usepackage{multirow}%
\usepackage{amsmath,amssymb,amsfonts}%
\usepackage{amsthm}%
\usepackage{mathrsfs}%
\usepackage[title]{appendix}%
\usepackage{xcolor}%
\usepackage{textcomp}%
\usepackage{manyfoot}%
\usepackage{booktabs}%
\usepackage{algorithm}%
\usepackage{algorithmicx}%
\usepackage{algpseudocode}%
\usepackage{listings}%
\usepackage{booktabs}
\usepackage{multirow}
\usepackage{makecell}
\theoremstyle{thmstyleone}%
\theoremstyle{thmstyletwo}%
\newcommand{\xmark}{\ensuremath{\times}}
\theoremstyle{thmstylethree}%

\begin{document}

\title[Article Title]{Patient-Agnostic Synthetic Pretraining for Efficient Patient-Specific Intraoperative 2D/3D Registration}


\author[1,2]{\fnm{Minheng} \sur{Chen}}\email{mxc2442@mavs.uta.edu}


\author*[2]{\fnm{Youyong} \sur{Kong}}\email{kongyouyong@seu.edu.cn}

\affil[1]{\orgdiv{Department of Computer Science and Engineering}, \orgname{ University of Texas at Arlington}, \orgaddress{\city{Arlington}, \postcode{76013}, \state{TX}, \country{USA}}}
\affil*[2]{\orgdiv{School of Computer Science and Engineering}, \orgname{Southeast University}, \orgaddress{\city{Nanjing}, \postcode{210096}, \state{Jiangsu}, \country{China}}}
% \affil[3]{\orgdiv{Department}, \orgname{Organization}, \orgaddress{\street{Street}, \city{City}, \postcode{610101}, \state{State}, \country{Country}}}


\abstract{
Intraoperative 2D/3D registration aligns preoperative CT volumes with intraoperative X-ray or fluoroscopic images and is essential for image-guided interventions. Recent learning-based and differentiable registration methods have shown promising accuracy, especially in patient-specific settings where abundant digitally reconstructed radiographs (DRRs) can be synthesized from the target CT. However, training a separate patient-specific model from scratch for every new patient is computationally inefficient and limits practical deployment. In this work, we propose an efficient patient-specific 2D/3D registration framework based on patient-agnostic synthetic pretraining and spherical similarity learning. The model is first pretrained on synthetic DRRs generated from multiple CT volumes to learn transferable pose-sensitive representations, and is then adapted to a new patient using only a limited number of synthetic projections from the target CT. To improve synthetic-to-real robustness without requiring anatomical labels, we introduce a segmentation-free domain randomization strategy that perturbs image intensity, projection physics, field-of-view, occlusion, and fluoroscopic artifacts. The adapted model provides an initial pose estimate, which is further refined using spherical similarity learning and differentiable Levenberg-Marquardt optimization. Experiments on multiple anatomical datasets evaluate whether patient-agnostic synthetic pretraining can improve the efficiency of patient-specific registration, with particular focus on the trade-off between adaptation cost and registration accuracy. The results demonstrate that patient-agnostic synthetic pretraining can significantly reduce patient-specific training requirements while preserving accurate intraoperative 2D/3D registration.
}

% \keywords{
%  \and
% intraoperative image guidance \and
% synthetic pretraining \and
% patient-specific adaptation \and
% domain randomization \and
% spherical similarity learning
% }

\maketitle
\section{Introduction}
\label{sec:intro}

Intraoperative 2D/3D registration aims to align intraoperative radiographic images, such as X-ray or fluoroscopy, with preoperative 3D volumes, such as CT. 
It is essential for image-guided interventions, where accurate spatial correspondence is required for surgical navigation, planning, and instrument or implant localization. 
Single-view 2D/3D registration is clinically attractive because it can simplify the imaging workflow and reduce radiation exposure, but it is also highly challenging due to projection ambiguity, limited anatomical visibility, and sensitivity to initialization.

Conventional methods usually solve this problem by generating digitally reconstructed radiographs (DRRs) from CT volumes and optimizing the rigid transformation that maximizes image similarity between the DRR and the intraoperative X-ray~\cite{grupp2018patch,penney1998comparison,knaan2003effective,frysch2021novel,de20163d,markelj2012review,chen2025introducing,chen2024optimization}. 
Hand-crafted similarity metrics, such as normalized cross-correlation (NCC), mutual information (MI), and local feature descriptors, are commonly combined with derivative-free optimizers such as BOBYQA or CMA-ES~\cite{toews2017phantomless,powell2009bobyqa,hansen2001completely}. 
Although these methods do not require annotated training data, their similarity landscapes are often highly non-convex outside a small capture range, leading to local minima and long optimization times.

Learning-based methods have been proposed to improve registration efficiency and robustness. 
Regression-based approaches directly predict the pose from paired X-ray/DRR images or 3D--2D inputs~\cite{li2025automatic,miao2016cnn,leroy2023structuregnet,zhao2024automatic}, while landmark-based methods rely on anatomical keypoints followed by geometric pose estimation~\cite{esteban2019towards,grimm2021pose,brandstatter2024rigid,markova2022global}. 
However, direct pose regression can generalize poorly to unseen patients or imaging conditions, and landmark-based methods require reliable anatomical landmark annotations, which are difficult to obtain in clinical fluoroscopy. 
These limitations motivate self-supervised synthetic learning strategies, where DRRs generated from CT volumes provide pose-supervised training pairs without manual annotation~\cite{unberath2018deepdrr,jaganathan2023self,gopalakrishnan2024intraoperative}.
Recently, differentiable rendering and similarity learning have provided a promising alternative between conventional optimization and direct regression. 
Gao \textit{et al.}~\cite{gao2020generalizing,gao2023fully} proposed learning a deep similarity metric whose gradient approximates pose-space geodesic directions on SE(3), thereby improving the capture range of 2D/3D registration. 
Subsequent studies further explored correlation-driven feature decomposition and embedded feature similarity optimization~\cite{chen2024fully,chen2024embedded}. 
Nevertheless, two important challenges remain. 
First, most learned similarity methods represent pose-induced image changes in Euclidean feature spaces, although rigid transformations naturally lie on non-Euclidean manifolds. 
Second, high-accuracy patient-specific registration often requires training a separate model from scratch for each new patient, which is computationally expensive and limits clinical scalability.

Patient-specific and patient-agnostic training represent different trade-offs. 
Patient-specific methods can synthesize abundant DRRs from the target CT and achieve high accuracy~\cite{zhang2023patient,gopalakrishnan2024intraoperative}, but require costly training for every new patient. 
Patient-agnostic models can be trained once across multiple subjects and reused on unseen patients~\cite{gao2020generalizing,chen2024fully}, but their accuracy is often limited by anatomical variability and the synthetic-to-real domain gap. 
This raises a natural question: 
\textit{can patient-agnostic synthetic pretraining provide an effective initialization for efficient patient-specific 2D/3D registration?}

In this work, we propose an efficient patient-specific 2D/3D registration framework based on patient-agnostic synthetic pretraining and spherical similarity learning. 
Instead of training a full patient-specific model from scratch, we first pretrain the registration network using synthetic DRRs generated from multiple CT volumes, and then adapt the pretrained model to a new patient using a limited number of synthetic projections from the target CT. 
To improve synthetic-to-real robustness without requiring anatomical labels, we introduce a segmentation-free domain randomization strategy that perturbs image appearance, projection geometry, field-of-view truncation, occlusion, and fluoroscopic artifacts.

Building on our previous spherical similarity learning framework~\cite{chen2026intraoperative}, image features are projected onto a hyperspherical manifold to provide a geometrically more consistent similarity measure. 
We further use a bi-invariant SO(4) pose-gradient formulation and differentiable Levenberg--Marquardt optimization for iterative pose refinement. 
Inspired by recent advances in parameter-efficient fine-tuning of vision foundation models, we systematically evaluate different adaptation settings, including the amount of patient-specific synthetic data and the model components selected for fine-tuning, to analyze the trade-off between registration accuracy and adaptation cost.
The main contributions of this work are summarized as follows:
\begin{itemize}
    \item We propose a patient-agnostic synthetic pretraining and patient-specific adaptation strategy for efficient intraoperative 2D/3D registration, reducing the need to train a separate model from scratch for every patient.
    
    \item We introduce a segmentation-free domain randomization scheme for synthetic pretraining, covering image appearance, projection geometry, field-of-view variation, occlusion, and fluoroscopic artifacts without requiring CT or X-ray segmentation.
    
    \item We build upon spherical similarity learning and an SO(4)-based pose-gradient formulation to provide a geometrically consistent similarity landscape for differentiable pose refinement.
    
    \item We systematically evaluate the trade-off between patient-specific adaptation cost and registration accuracy, including adaptation data size, trainable modules, training time, and the effect of domain randomization.
\end{itemize}
\begin{figure*}[h!]
\centering
\includegraphics[width=\linewidth]{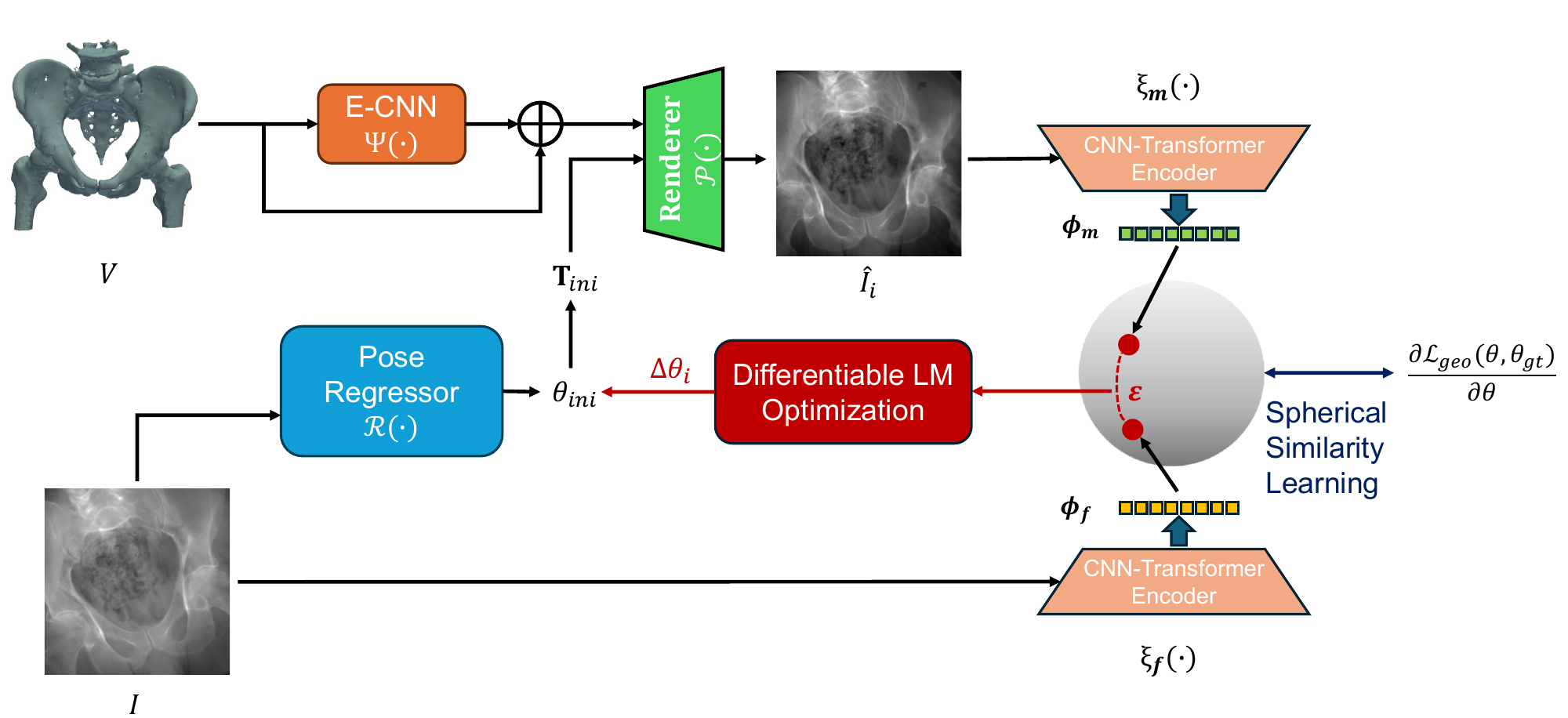}
\caption{
    \textbf{Overview of the proposed framework.}
    We first employ a regressor to initialize the pose and then refine it using differentiable Levenberg-Marquardt optimization based on spherical similarity learning. Spherical similarity learning consists of two main components: extracting image feature representations using CNN-Transformer encoders $\xi(\cdot)$ and projecting these embeddings into hypersphere space, where the geodesic distance between them is computed as a measure of deep similarity. During training, we enforce the gradient of this deep similarity with respect to $\theta$ to approximate the gradient of the geodesic distance between $\theta$ and the ground truth $\theta_{gt}$ in SE(3).
}
\label{overview}
\end{figure*}
\section{Related Work}
\label{sec:related}

\subsection{Learning-based 2D/3D Registration}

Learning-based methods have been widely studied for 2D/3D registration because they can improve computational efficiency and enlarge the capture range compared with purely optimization-based pipelines. Existing methods can be broadly grouped into direct regression, landmark-based estimation, and hybrid two-stage frameworks.
Regression-based methods directly predict the 3D pose from a 2D radiograph, a 3D volume, or paired 2D/3D inputs~\cite{guo2021end,leroy2023structuregnet,miao2016real,zhao2024automatic}. These methods enable fast feed-forward pose estimation, which is attractive for intraoperative navigation and robotic-assisted interventions. However, direct regression often lacks the precision required for high-accuracy alignment, especially under single-view ambiguity, appearance variation, or synthetic-to-real domain shift.
Landmark-based methods estimate the transformation from anatomical landmarks, local descriptors, or point-to-plane correspondences~\cite{esteban2019towards,grimm2021pose,liao2019multiview,shrestha2023x,shrestha2024rayemb}. They typically formulate registration as a geometric pose estimation problem, such as perspective-n-point (PnP)~\cite{li2012robust}, or use point-to-plane solvers for 2D/3D alignment~\cite{jaganathan2021deep,jaganathan2023self,schaffert2020learning,wang2017dynamic}. While these methods can be accurate when reliable correspondences are available, their performance depends strongly on the visibility and localization accuracy of anatomical structures~\cite{toews2017phantomless}. In clinical fluoroscopy, landmarks may be occluded, truncated, or degraded by low contrast and surgical instruments, often requiring robust estimation strategies such as RANSAC~\cite{fischler1981random}.

Hybrid two-stage methods combine learning-based initialization with iterative refinement~\cite{chen2024embedded,zhang2023patient,gopalakrishnan2024intraoperative,downs2025improving,gopalakrishnan2025rapid}. A neural network first predicts an initial pose, which is then refined using differentiable rendering, learned similarity optimization, or derivative-free search~\cite{gopalakrishnan2022fast,li2025automatic}. This strategy is effective in patient-specific scenarios, where synthetic projections from a fixed preoperative CT can provide dense training supervision. However, training a separate patient-specific model from scratch is computationally costly, whereas patient-agnostic models are reusable but usually less accurate due to inter-patient anatomical variability. This trade-off motivates the patient-agnostic pretraining and patient-specific adaptation strategy studied in this work.

\subsection{Synthetic Learning and Domain Randomization}

A major challenge in learning-based 2D/3D registration is the lack of real X-ray/CT pairs with accurate ground-truth poses. 
Synthetic learning addresses this limitation by generating DRRs from CT volumes under known projection geometries, thereby providing pose-supervised training data without manual annotation~\cite{unberath2018deepdrr,gopalakrishnan2024intraoperative,toth2019training}. 
Since pose labels are directly determined by the rendering process, DRR-based learning is particularly suitable for 2D/3D registration.
Synthetic data have also become an important strategy for improving generalization in learning-based medical imaging~\cite{gopinath2024synthetic,dey2025learning,fu2025synthesizing}. 
Representative Synth-based methods, including SynthSeg~\cite{billot2023synthseg,billot2023robust,laso2024quantifying}, SynthStrip~\cite{hoopes2022synthstrip}, and SynthMorph~\cite{hoffmann2022synthmorph,hoffmann2024synthmorph}, use randomized synthetic images to generalize across imaging protocols, contrasts, resolutions, and populations. 
These studies suggest that carefully designed synthetic distributions can help models learn task-relevant anatomical and geometric representations that are less dependent on a specific acquisition setting.

For intraoperative 2D/3D registration, however, models must bridge the gap between synthetic DRRs and real fluoroscopic images~\cite{unberath2019enabling,gao2023synthetic}. 
DRRs generated from simplified projection models may not reproduce scatter, detector response, truncation, surgical tools, or patient-positioning effects. 
Domain randomization has therefore been used to diversify synthetic training data and improve robustness to real imaging conditions~\cite{huang2024pele,zakharov2022photo}. 
However, perturbations in 2D/3D registration must be carefully controlled, since they should improve appearance invariance without destroying pose-sensitive anatomical information.
Synthetic DRR learning can be applied in both patient-agnostic and patient-specific settings. 
Patient-agnostic models are reusable across unseen patients but are often limited by anatomical variability and the synthetic-to-real gap. 
Patient-specific models can learn more accurate anatomy-dependent representations from the target CT~\cite{gopalakrishnan2024intraoperative,gopalakrishnan2025rapid}, but training or adapting a separate model for each patient introduces additional computational cost.

In this work, we study whether patient-agnostic synthetic pretraining can reduce the cost of patient-specific 2D/3D registration. 
We first learn transferable pose-sensitive representations from DRRs generated across multiple CT volumes, and then adapt the pretrained model to a new patient using limited synthetic projections from the target CT. 
Unlike segmentation-driven synthetic frameworks such as SynthSeg or SynthMorph, our setting requires neither anatomical labels nor segmentation masks, making the proposed strategy annotation-free and segmentation-free.

\subsection{Similarity Learning for Image Registration}

Image registration relies critically on the choice of image similarity measure. 
Traditional metrics, such as sum of squared differences (SSD), normalized cross-correlation (NCC), and mutual information (MI), have been widely used in medical image registration~\cite{markelj2012review}. 
However, their similarity landscapes are often non-convex over a large pose range, which can trap iterative optimization in local minima. 
This issue is especially challenging in single-view 2D/3D registration, where projection ambiguity and the X-ray/DRR appearance gap are substantial.

Similarity learning addresses this problem by learning task-specific feature representations and similarity functions from data~\cite{chen2024fully,qin2019unsupervised,grzech2022variational}. 
Learned metrics can provide more informative gradients for optimization and improve robustness to local appearance variations~\cite{ronchetti2023disa,grzech2024unsupervised,sideri2023mad,mok2024modality}. 
In 2D/3D registration, Gu \textit{et al.}~\cite{gu2020extended} introduced similarity learning through Riemannian pose-gradient estimation, and Gao \textit{et al.}~\cite{gao2020generalizing,gao2023fully} extended this idea into a fully differentiable framework. 
Subsequent work further improved learned similarity optimization through correlation-driven feature decomposition and parameter-specific initialization~\cite{chen2024fully,chen2024embedded}.
Most existing similarity learning frameworks represent image features in Euclidean latent spaces, although rigid transformations and pose-induced image changes naturally lie on non-Euclidean manifolds. 
To provide a more geometrically consistent similarity landscape, our method builds on spherical similarity learning and a bi-invariant SO(4) pose-gradient formulation. 
Combined with patient-agnostic synthetic pretraining and patient-specific adaptation, this learned similarity metric can be efficiently transferred to new patients while maintaining sensitivity to anatomical misalignment.
\section{Method}
\label{sec:method}
\begin{figure*}[h!]
\centering
\includegraphics[width=\linewidth]{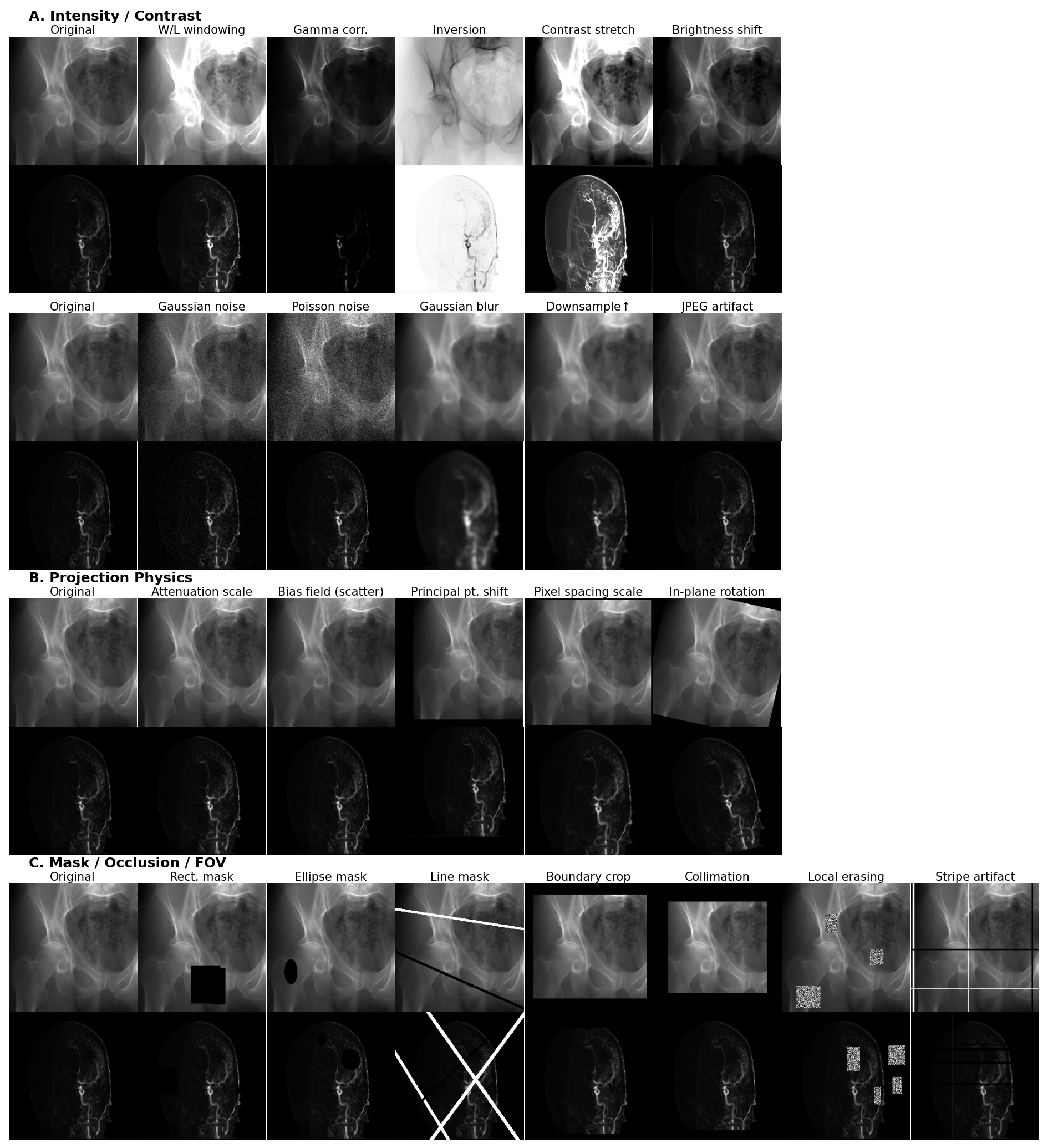}
\caption{Illustration of the proposed segmentation-free domain randomization. 
The augmentations are grouped into intensity/contrast perturbations, projection-physics variations, and mask \&occlusion \& FOV transformations, covering common synthetic-to-real differences between DRRs and intraoperative X-ray images.
}
\label{fig::randomization}
\end{figure*}

\subsection{Problem Formulation}

Given a preoperative CT volume \(V\) and an intraoperative X-ray image \(I^{x}\), 2D/3D registration aims to estimate the rigid transformation that aligns the CT volume with the X-ray image. 
Let \(\mathcal{P}(\cdot)\) denote a differentiable projector that generates a digitally reconstructed radiograph (DRR) from the CT volume under a given pose. 
The rigid transformation is represented as:
\begin{equation}
    \mathbf{T} =
    \begin{bmatrix}
        \mathbf{R} & \mathbf{t} \\
        0 & 1
    \end{bmatrix}
    \in SE(3)
\end{equation}
where \(\mathbf{R}\in SO(3)\) is the rotation matrix and \(\mathbf{t}\in \mathbb{R}^{3}\) is the translation vector. 
The registration objective is formulated as:
\begin{equation}
    \mathbf{T}^{*}
    =
    \arg\min_{\mathbf{T}\in SE(3)}
    \mathcal{S}
    \left(
    I^{x},
    \mathcal{P}(\mathbf{T})\circ V
    \right)
\end{equation}
where \(\mathcal{S}(\cdot,\cdot)\) measures the dissimilarity between the intraoperative X-ray image and the rendered DRR.
\subsection{Network Architecture}

The overall framework follows our conference version~\cite{chen2026intraoperative} as shown in Fig.~\ref{overview}. 
Given an input X-ray image and a CT volume, a pose regressor \(\mathcal{R}(\cdot)\) first predicts an initial pose parameter \(\theta_{\mathrm{ini}}\in\mathfrak{se}(3)\). 
The predicted parameter is mapped to \(SE(3)\) through the exponential map:
\begin{equation}
    \mathbf{T}_{\mathrm{ini}}
    =
    \mathrm{Exp}(\theta_{\mathrm{ini}})
\end{equation}
The initial transformation is then used to render an initial DRR from the CT volume.

To improve the representation of the projected volume, we employ an equivariant CNN module~\cite{billot2024se,moyer2021equivariant} \(\Psi(\cdot)\) to extract transformation-aware 3D features. 
The enhanced volume is obtained by residual addition:
\begin{equation}
    \hat{V}=V+\Psi(V),
\end{equation}
and the corresponding moving DRR is generated as:
\begin{equation}
    \hat{I}_{m}(\mathbf{T})
    =
    \mathcal{P}(\mathbf{T})\circ \hat{V}
\end{equation}
The fixed image and moving DRR are then encoded by two feature extractors, \(\xi_f\) and \(\xi_m\), which share the same architecture:
\begin{equation}
    \phi_f = \xi_f(I_f),
    \quad
    \phi_m(\mathbf{T}) = \xi_m(\hat{I}_{m}(\mathbf{T}))
\end{equation}
The learned feature maps are used to compute the spherical similarity described below.
\subsection{Spherical Similarity Learning}

To obtain a geometrically consistent similarity metric, we embed the learned feature vectors onto a hyperspherical manifold. 
Given feature maps \(\phi_f\) and \(\phi_m\), we apply a spherical exponential mapping along the channel dimension:
\begin{equation}
    \Phi_f=\mathrm{EXP}(\phi_f),
    \quad
    \Phi_m=\mathrm{EXP}(\phi_m)
\end{equation}
The spherical discrepancy between the fixed and moving images is defined as:
\begin{equation}
\label{eq:similarity}
    \varepsilon(\mathbf{T})
    =
    \sum_{u}
    \left(
    1-\Phi_m(u,\mathbf{T})^{\top}\Phi_f(u)
    \right)
\end{equation}
where \(u\) indexes spatial feature locations. 
A smaller value of \(\varepsilon(\mathbf{T})\) indicates better alignment between the fixed image and the moving DRR.

Following our previous work, the similarity network is trained using a double-backward strategy. 
Instead of directly supervising the absolute similarity value, we align the gradient of the learned similarity with the gradient of a pose-space geodesic distance:
\begin{equation}
    \nabla_{\theta}\mathcal{L}_{\mathrm{net}}(\theta)
    \approx
    \nabla_{\theta}\mathcal{L}_{\mathrm{geo}}(\theta,\theta_{\mathrm{gt}})
\end{equation}
where \(\mathcal{L}_{\mathrm{net}}=\varepsilon(\mathbf{T})\). 
To provide more geometrically consistent supervision, the pose is embedded into a bi-invariant \(SO(4)\) representation, and the geodesic discrepancy is computed in this space as in our conference version~\cite{chen2026intraoperative}.
\subsection{Segmentation-Free Domain Randomization}

To improve the transferability of synthetic DRRs to real intraoperative X-ray images, we introduce a segmentation-free domain randomization strategy. 
Unlike segmentation-based synthetic frameworks, the proposed augmentation does not require CT segmentation, X-ray segmentation, anatomical labels, or landmark masks. 
It operates directly on DRR images and contains three operation pools: intensity and contrast randomization \(\mathcal{G}_{I}\), projection-physics randomization \(\mathcal{G}_{P}\), and mask, occlusion, and field-of-view randomization \(\mathcal{G}_{M}\).
The operations used for randomization are illustrated in Fig.~\ref{fig::randomization}.
The intensity and contrast pool \(\mathcal{G}_{I}\) simulates photometric differences between DRRs and real X-ray images, including window-level adjustment, gamma correction, intensity inversion, contrast stretching, brightness shifts, Gaussian and Poisson noise, Gaussian blur, downsampling-upsampling, and JPEG compression. 
The projection-physics pool \(\mathcal{G}_{P}\) approximates acquisition-related variations, including attenuation scaling, smooth multiplicative bias fields, principal-point shifts, pixel-spacing or zoom changes, and in-plane detector rotations. 
The mask, occlusion, and field-of-view pool \(\mathcal{G}_{M}\) simulates content mismatch commonly observed in fluoroscopy, including rectangular and elliptical masks, line-shaped occlusions, boundary cropping, collimation masks, local erasing, and stripe artifacts.

Let \(\mathcal{A}(\cdot)\) denote the stochastic augmentation operator. 
During synthetic training, the fixed DRR generated from the ground-truth pose is transformed into a pseudo-X-ray image:
\begin{equation}
    \tilde{I}_{\mathrm{gt}}
    =
    \mathcal{A}
    \left(
    \mathcal{P}(\mathbf{T}_{\mathrm{gt}})\circ V
    \right)
\end{equation}
The moving DRR remains differentiably rendered from the perturbed pose:
\begin{equation}
    I_m(\mathbf{T})
    =
    \mathcal{P}(\mathbf{T})\circ V
\end{equation}
This design encourages robustness to appearance variation while preserving differentiability with respect to pose parameters.
For each operation pool \(g\in\{I,P,M\}\), two distinct operations are sampled without replacement and applied in random order:
\begin{equation}
    \mathcal{A}_{g}^{*}
    =
    A_{g}^{(2)} \circ A_{g}^{(1)},
    \quad
    A_{g}^{(1)},A_{g}^{(2)}\in\mathcal{G}_{g}
\end{equation}
The final compound augmentation is defined as
\begin{equation}
    \mathcal{A}
    =
    \left(\mathcal{A}_{M}^{*}\right)^{z_M}
    \circ
    \left(\mathcal{A}_{P}^{*}\right)^{z_P}
    \circ
    \left(\mathcal{A}_{I}^{*}\right)^{z_I}
\end{equation}
where \(z_g\sim \mathrm{Bernoulli}(p_g)\), and \(p_I=p_P=p_M=0.5\). 
When \(z_g=0\), the corresponding operation pool is replaced by the identity mapping.
\subsection{Patient-Agnostic Pretraining and Patient-Specific Adaptation}
\begin{figure}[h!]
\centering
\includegraphics[width=\linewidth]{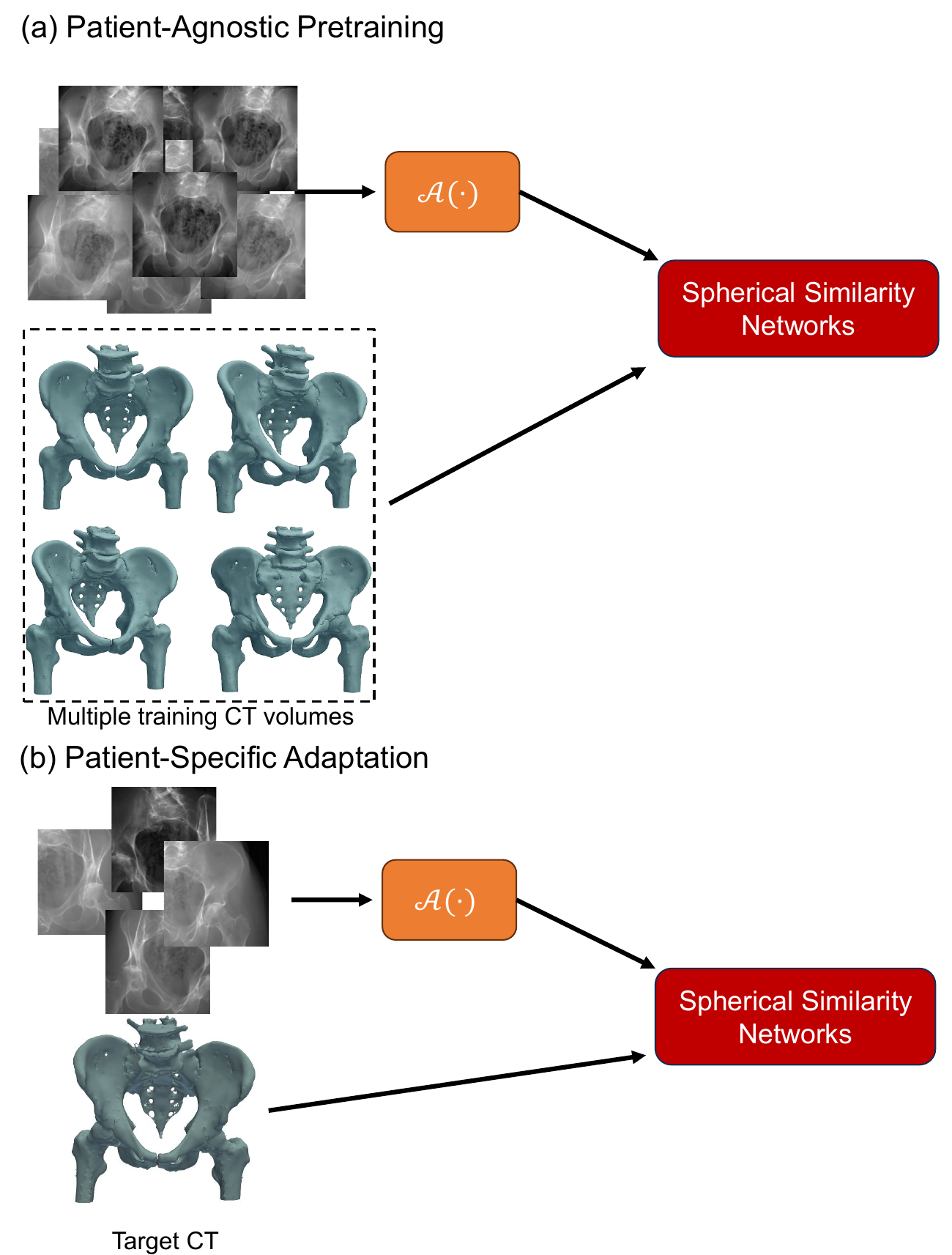}
\caption{Overview of the proposed pretraining--adaptation strategy. 
(a) Patient-agnostic pretraining learns spherical similarity networks from synthetic DRRs generated from multiple training CT volumes with segmentation-free domain randomization \(\mathcal{A}(\cdot)\). 
(b) Patient-specific adaptation fine-tunes the pretrained model using limited randomized synthetic projections from the target CT.
}
\label{fig::fig1}
\end{figure}
\begin{figure}[h!] \centering \includegraphics[width=\linewidth]{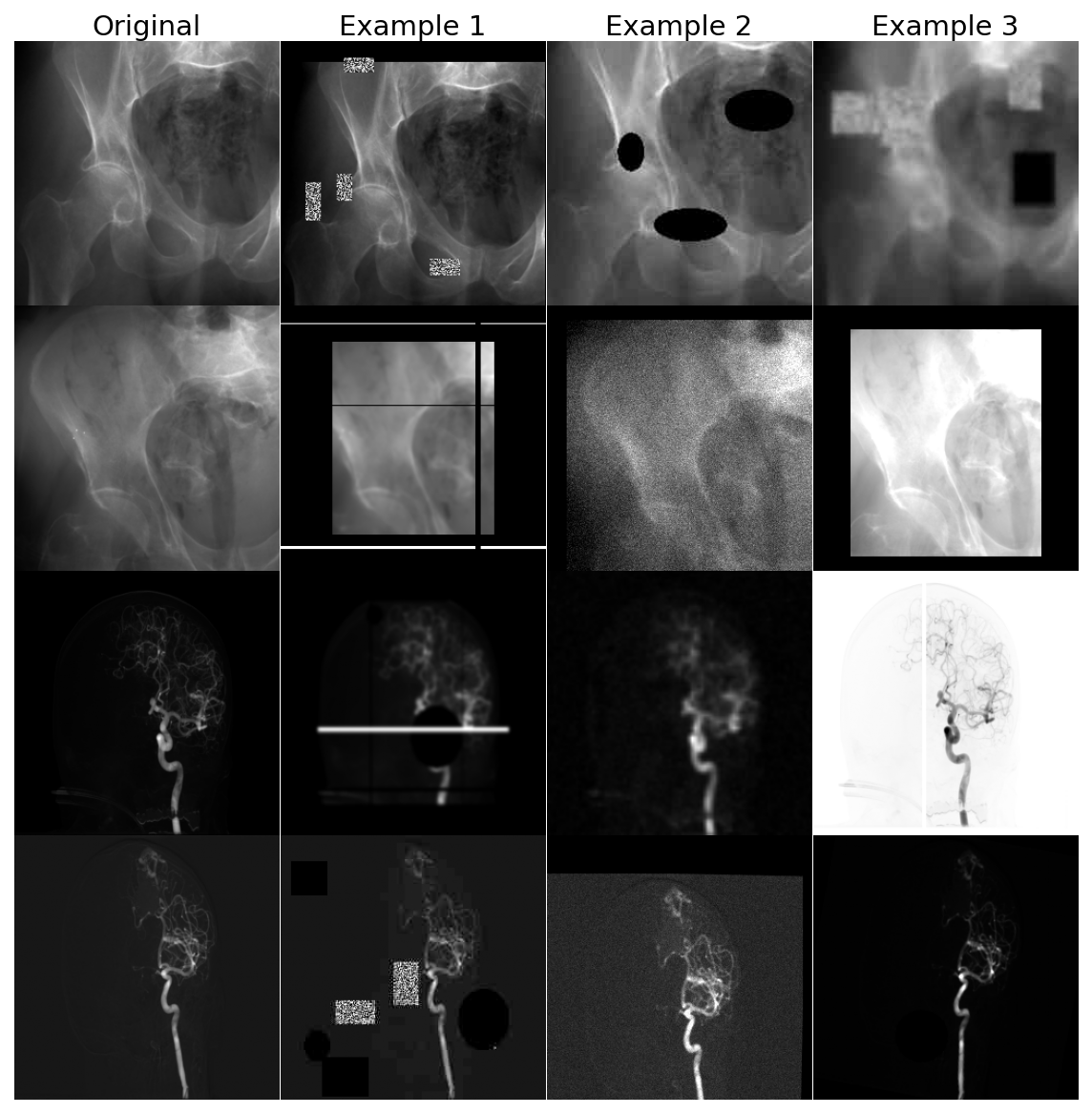} \caption{Examples of segmentation-free domain randomization. The first column shows the original projection, and the remaining columns show randomly augmented examples with intensity, physics, occlusion, and FOV perturbations. } \label{fig::examples} \end{figure}
The proposed training strategy consists of patient-agnostic synthetic pretraining followed by patient-specific adaptation. 
In the pretraining stage, synthetic DRR pairs are generated from multiple training CT volumes. 
For each sample, a ground-truth pose \(\mathbf{T}_{\mathrm{gt}}\) is sampled within an anatomy-specific pose range, and a perturbed pose \(\mathbf{T}\) is sampled around it to generate the moving DRR. 
The fixed image is randomized by \(\mathcal{A}(\cdot)\), while the moving image is rendered from the perturbed pose.
Given a new patient, the pretrained model is fine-tuned using a limited number of synthetic DRRs generated only from the target CT volume \(V_{\mathrm{new}}\). 
The adaptation set is written as:
\begin{equation}
    \mathcal{D}_{\mathrm{adapt}}
    =
    \left\{
    \left(
    \tilde{I}_{\mathrm{gt}}^{k},
    I_m^{k},
    \mathbf{T}^{k},
    \mathbf{T}_{\mathrm{gt}}^{k}
    \right)
    \right\}_{k=1}^{K}
\end{equation}
where \(K\) is the number of patient-specific synthetic samples. 
During adaptation, we use a modified augmentation operator \(\mathcal{A}_{\mathrm{adapt}}\), which follows the same compound augmentation strategy but excludes the mask, occlusion, and field-of-view pool \(\mathcal{G}_{M}\). 
This avoids artificial removal of patient-specific anatomical structures while retaining mild appearance and acquisition variability.
Compared with patient-specific training from random initialization, the proposed strategy starts from a patient-agnostic pretrained representation and therefore requires fewer synthetic samples and fewer training iterations. 
Motivated by recent advances in parameter-efficient fine-tuning, we evaluate different adaptation settings by varying both the amount of patient-specific synthetic data and the set of trainable modules.

\begin{table*}[h!]
    \centering
    \caption{
    Comparison of the proposed pretrained fine-tuning method with existing patient-specific 2D/3D registration baselines on DeepFluoro and Ljubljana.
    Our pretrained FT method adapts a patient-agnostic pretrained model to each target patient using synthetic projections from the target CT.
    SMSR denotes the sub-millimeter success rate, defined as the percentage of test cases with mTRE \(<1\) mm.
    Median, 75th percentile, and 95th percentile mTREs are reported.
    For accuracy metrics, the best results are \textbf{bolded}, and the second-best results are \underline{underlined}.
    Run time is reported for reference and is not used for ranking.
    }
    \label{tab::patient-specific}
    \resizebox{\textwidth}{!}{
    \begin{tabular}{c l c c c c c}
        \toprule
        \multirow{2}{*}{\textbf{Dataset}} 
        & \multirow{2}{*}{\textbf{Method}} 
        & \multirow{2}{*}{\textbf{SMSR}} 
        & \textbf{Median} 
        & \multicolumn{2}{c}{\textbf{Percentile (mm)}} 
        & \multirow{2}{*}{\textbf{Run Time}} \\
        \cmidrule(lr){5-6}
        & & & \textbf{(mm)} & \textbf{75\%} & \textbf{95\%} & \\
        \midrule

        \multirow{7}{*}{\textbf{DeepFluoro}} 
        & PSSS-reg~\cite{zhang2023patient}         
        & 56.0\%  & 0.93  & 2.51  & 5.57  & 12.7 s \\
        & PoseNet~\cite{bui2017x}       
        & 4.3\%   & 16.6  & 22.0  & 29.2  & 0.1 s \\
        & DFLNet~\cite{grupp2020automatic}      
        & 36.6\%  & 3.20  & 7.29  & 13.1  & 1.0 s \\
        & SCR-reg~\cite{shrestha2023x}      
        & 33.3\%  & 4.70  & 9.59  & 12.8  & 1.1 s \\
        & DiffPose~\cite{gopalakrishnan2024intraoperative}     
        & 83.1\%  & 0.60  & \underline{0.89}  & \underline{1.47}  & 5.3 s \\
        \cmidrule{2-7}
        & Ours (from scratch)   
        & \textbf{86.1\%} & \textbf{0.51} & \textbf{0.85} & \textbf{1.42} & 6.2 s \\
        & Ours (pretrained FT)   
        & \underline{84.7\%} & \underline{0.55} & \underline{0.89} & 1.58 & 6.3 s \\

        \midrule

        \multirow{5}{*}{\textbf{Ljubljana}} 
        & PSSS-reg~\cite{zhang2023patient}         
        & 40.0\%  & 2.48  & 5.87  & 11.3  & 15.3 s \\
        & PoseNet~\cite{bui2017x}       
        & 0\%     & 23.3  & 26.2  & 29.2  & \(<0.1\) s \\
        & DiffPose~\cite{gopalakrishnan2024intraoperative}     
        & \underline{80.0\%}  & 0.63  & \underline{0.94}  & 1.78  & 6.0 s \\
        \cmidrule{2-7}
        & Ours (from scratch)  
        & \textbf{85.0\%} & \underline{0.55} & \textbf{0.85} & \underline{1.35} & 6.5 s \\
        & Ours (pretrained FT)    
        & \textbf{85.0\%} & \textbf{0.52} & \textbf{0.85} & \textbf{1.33} & 6.4 s \\

        \bottomrule
    \end{tabular}
    }
\end{table*}
\subsection{Training and Inference}

As shown in Fig.~\ref{fig::fig1}, the overall training strategy consists of two stages: patient-agnostic synthetic pretraining and patient-specific adaptation. 
In the first stage, the model is pretrained using synthetic DRRs generated from multiple training CT volumes. 
For each training sample, a ground-truth pose is sampled within the anatomy-specific pose range, and a perturbed pose is used to generate the moving DRR. 
The pose regressor is optimized by minimizing the geodesic discrepancy between the predicted initial pose and the ground-truth pose, while the spherical similarity network is trained using the double-backward gradient alignment strategy described above.
In the second stage, the pretrained model is adapted to a new patient using a limited number of synthetic DRRs generated only from the target CT volume. 
The same pose-supervised training objectives are used during adaptation, but only selected modules are fine-tuned depending on the adaptation setting. 
This strategy transfers the patient-agnostic pose-sensitive representation to the target anatomy while avoiding training the entire patient-specific model from random initialization.

During inference, the adapted regressor first predicts an initial pose \(\mathbf{T}_{\mathrm{ini}}\). 
The pose is then refined by minimizing the learned spherical discrepancy:
\begin{equation}
    \mathbf{T}^{*}
    =
    \arg\min_{\mathbf{T}}
    \varepsilon(\mathbf{T})
\end{equation}
We use differentiable Levenberg-Marquardt optimization for iterative refinement. 
At iteration \(i\), the pose increment is computed as:
\begin{equation}
    \Delta \theta_i
    =
    \left(
    J_i^{\top}WJ_i+\lambda I
    \right)^{-1}
    J_i^{\top}W\mathbf{r}(\theta_{i-1})
\end{equation}
where \(J_i\) is the Jacobian of the residual vector \(\mathbf{r}(\theta_{i-1})\), \(W\) is the weighting matrix, and \(\lambda\) is the damping parameter. 
The pose is updated by left multiplication:
\begin{equation}
    \mathbf{T}_{i}
    =
    \mathrm{Exp}(\Delta \theta_i)\mathbf{T}_{i-1}
\end{equation}
The final pose after LM refinement is used as the registration result.
\section{Experiments and Results}
\begin{table*}[h!]
\centering
\caption{
Main comparison of patient-specific adaptation strategies on DeepFluoro and Ljubljana.
Speed-up is computed relative to patient-specific training from scratch on each dataset.
}
\label{tab:main_adaptation}
\resizebox{\textwidth}{!}{
\begin{tabular}{llcccccc}
\toprule
Dataset 
& Strategy
& Pretraining
& \makecell{Fine-tuning\\amount}
& \makecell{Training\\time (h)}
& \makecell{Speed-up\\vs. scratch}
& SMSR $\uparrow$
& \makecell{Median / 95\%\\mTRE (mm) $\downarrow$} \\
\midrule

\multirow{6}{*}{DeepFluoro}
& From scratch
& No  & Full & 20.70 & 1.0$\times$  & 86.1\% & 0.51 / 1.42 \\
& Patient-agnostic only
& Yes & 0    & 0     & --           & 19.6\% & 4.50 / 22.63 \\
& Pretrain + 10\% FT
& Yes & 10\% & 0.04  & 517.5$\times$ & 42.4\% & 2.78 / 15.86 \\
& Pretrain + 25\% FT
& Yes & 25\% & 0.10  & 207.0$\times$ & 60.9\% & 0.92 / 8.94 \\
& Pretrain + 50\% FT
& Yes & 50\% & 0.22  & 94.1$\times$  & 74.3\% & 0.75 / 1.67 \\
& Pretrain + full FT
& Yes & Full & 0.47  & 44.0$\times$  & 84.7\% & 0.55 / 1.58 \\
\midrule

\multirow{6}{*}{Ljubljana}
& From scratch
& No  & Full & 20.90 & 1.0$\times$  & 85.0\% & 0.55 / 1.35 \\
& Patient-agnostic only
& Yes & 0    & 0     & --           & 23.4\% & 4.26 / 19.33 \\
& Pretrain + 10\% FT
& Yes & 10\% & 0.06  & 348.3$\times$ & 70.0\% & 0.81 / 2.73 \\
& Pretrain + 25\% FT
& Yes & 25\% & 0.13  & 160.8$\times$ & 80.0\% & 0.63 / 1.86 \\
& Pretrain + 50\% FT
& Yes & 50\% & 0.25  & 83.6$\times$  & 81.0\% & 0.59 / 1.59 \\
& Pretrain + full FT
& Yes & Full & 0.48  & 43.5$\times$  & 85.0\% & 0.52 / 1.33 \\

\bottomrule
\end{tabular}
}
\end{table*}

\subsection{Datasets and Evaluation Metrics}
\label{sec:datasets_metrics}

We evaluate the proposed method on two publicly available datasets covering different anatomical regions and imaging scenarios.

\begin{itemize}
    \item \textbf{DeepFluoro}: The DeepFluoro dataset contains six pelvic CT scans and 366 real fluoroscopic X-ray images~\cite{grupp2020automatic}. The dataset provides the calibrated intrinsic matrix of the C-arm imaging system and the ground-truth extrinsic matrix for each X-ray image. In addition, 14 manually annotated anatomical landmarks are available for each CT scan, enabling quantitative evaluation of target registration error.

    \item \textbf{Ljubljana}: The Ljubljana dataset~\cite{pernus20133d} contains ten clinical 3D cone-beam computed tomography (CBCT) subtracted angiography scans from patients undergoing cerebral endovascular treatment. Each CBCT scan is associated with two X-ray images, and calibrated intrinsic and extrinsic matrices are provided for registration evaluation.
\end{itemize}

\begin{figure*}[h!] 
\centering \includegraphics[width=\linewidth]{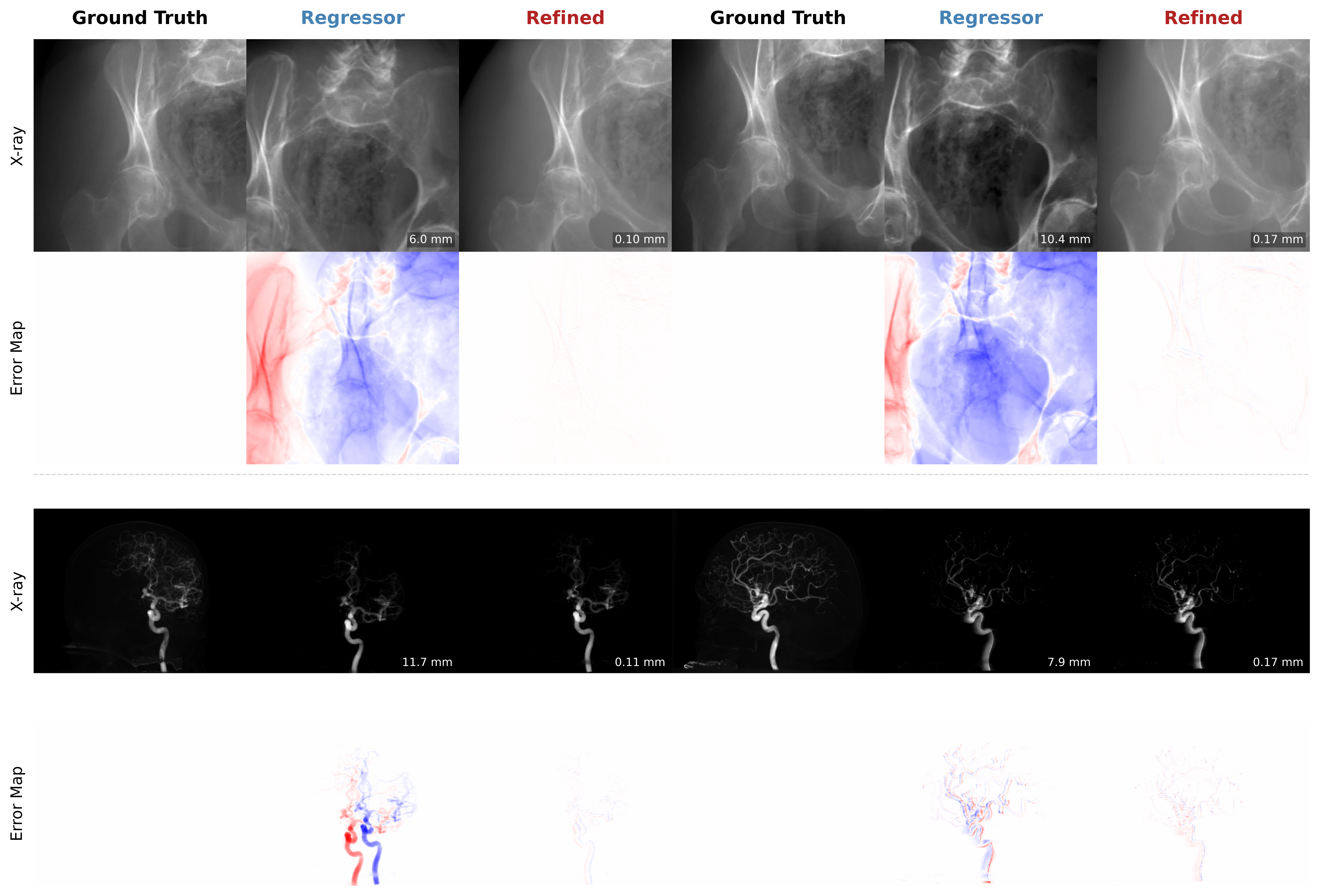} 
\caption{
Qualitative registration results on DeepFluoro and Ljubljana. 
For each case, we show the ground-truth X-ray, the DRR generated from the regressor initialization, and the DRR after differentiable LM refinement. 
The corresponding error maps show that the refinement step substantially reduces the residual misalignment from the initial pose estimate.
}
\label{fig::qualitative} \end{figure*}
\begin{figure*}[h!] \centering \includegraphics[width=\linewidth]{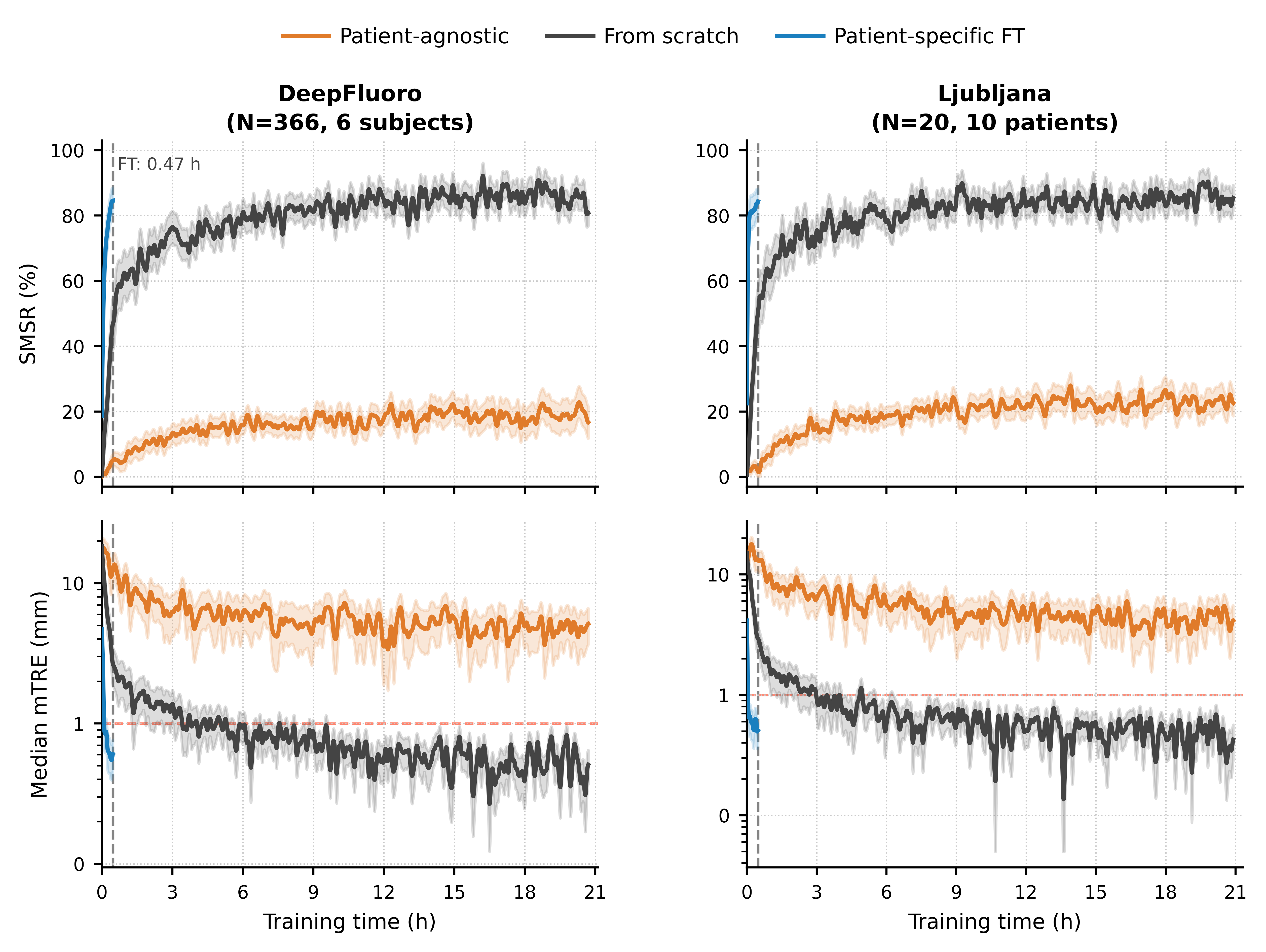} \caption{
Training efficiency comparison between patient-specific training from scratch and patient-agnostic pretraining with patient-specific fine-tuning. 
Curves show SMSR and median mTRE as a function of training time on DeepFluoro and Ljubljana. 
Patient-agnostic direct transfer provides limited accuracy, whereas patient-specific fine-tuning rapidly reaches near from-scratch performance within approximately 0.5 h. 
The red dashed line indicates the 1 mm mTRE threshold for sub-millimeter registration.
} \label{fig::training} \end{figure*}

Following prior studies~\cite{gopalakrishnan2024intraoperative,unberath2018deepdrr,momeni2024voxel}, we preprocess the generated DRRs to reduce the intensity discrepancy between simulated projections and real X-ray images. X-ray imaging records the attenuation of photons after traversing tissue, whereas DRR rendering based on line integration often represents accumulated absorption along each ray. As a result, high-density structures may appear with opposite contrast in DRRs and real X-ray images. To make the grayscale distribution of DRRs generated by Siddon's ray-tracing method~\cite{siddon1985fast} more consistent with real X-ray images, we apply the following intensity inversion:
\begin{equation}
    \tilde{I}
    =
    1 -
    \frac{\log(1+I)}{\log(1+I_0)}
\end{equation}
where \(I\) denotes the rendered DRR intensity and \(I_0\) is the maximum intensity used for normalization. The logarithmic transformation improves numerical stability and enhances contrast consistency between DRRs and X-ray images.
The evaluation metrics used in our experiments are as follows:

\begin{itemize}
    \item \textbf{Mean target registration error (mTRE)}: mTRE measures the average Euclidean distance between corresponding target points transformed by the estimated pose and the ground-truth pose. For DeepFluoro, we use the provided anatomical landmarks. For Ljubljana, where manual landmarks are not available, we compute the error using sampled 3D points from the reconstructed anatomical structure, following the calibrated ground-truth projection geometry. We report the median, 75th percentile, and 95th percentile mTRE.
    \item \textbf{Sub-millimeter success rate (SMSR)}: SMSR is defined as the percentage of test cases with mTRE below 1 mm. This metric reflects the proportion of registrations that achieve sub-millimeter accuracy, which is important for high-precision image-guided interventions.

    \item \textbf{Training time}: Since the main goal of this work is to reduce patient-specific adaptation cost, we also report the time required for patient-specific fine-tuning under different adaptation settings.
\end{itemize}

\subsection{Training and Implementation Details}
\label{sec:implementation}

All models were implemented in PyTorch and trained on an NVIDIA A6000 GPU. 
For patient-specific evaluation, we adopted a leave-one-out protocol. 
For each target patient, the patient-agnostic model was pretrained using CT volumes from all remaining subjects, ensuring that the target CT was excluded from pretraining. 
The pretrained model was then adapted using synthetic DRRs generated only from the target patient's CT volume.
We follow the pose parameter setting in~\cite{gopalakrishnan2025rapid},
for both patient-agnostic pretraining and patient-specific adaptation, synthetic DRR pairs were generated by sampling ground-truth poses within anatomy-specific pose ranges and then perturbing them to obtain moving projections. 
For the DeepFluoro dataset, rotations were sampled from
\(\alpha\in[-45^\circ,45^\circ]\), \(\beta\in[-45^\circ,45^\circ]\), and \(\gamma\in[-15^\circ,15^\circ]\), while translations were sampled from
\(x\in[-150,150]\) mm, \(y\in[-1000,-450]\) mm, and \(z\in[-150,150]\) mm. 
For the Ljubljana dataset, rotations were sampled from
\(\alpha\in[-45^\circ,90^\circ]\), \(\beta\in[-5^\circ,5^\circ]\), and \(\gamma\in[-5^\circ,5^\circ]\), while translations were sampled from
\(x\in[-25,25]\) mm, \(y\in[700,800]\) mm, and \(z\in[-25,25]\) mm.

During patient-agnostic pretraining, the fixed projection was transformed into a pseudo-X-ray image using the proposed segmentation-free domain randomization, while the moving projection was rendered from the perturbed pose to preserve differentiability with respect to pose parameters~\cite{gopalakrishnan2022fast}. 
Unlike the patient-agnostic setting in our conference version~\cite{chen2026intraoperative}, where the pose regressor was replaced by RTPIv3~\cite{chen2024embedded} to incorporate CT spatial information and trained with 750k regressor samples and 300k similarity-learning samples, we keep the same regressor architecture as in the patient-specific setting in this study. 
This design provides a controlled comparison between patient-specific training from scratch and pretrained fine-tuning, ensuring that performance differences mainly reflect the effect of synthetic pretraining and adaptation rather than changes in the regressor architecture. 
For consistency in training-data scale across experiments, we also reduced the number of self-supervised training samples to 500k for the pose regressor and 200k for the similarity network.

We used the Adam optimizer with an initial learning rate of \(1\times10^{-3}\), weight decay of \(1\times10^{-3}\), and a cyclic learning-rate scheduler. 
For patient-specific adaptation, the pretrained model was fine-tuned using synthetic DRRs generated only from the target patient's CT volume. 
Unless otherwise specified, we used 20k patient-specific synthetic samples for adaptation and fine-tuned the pose regressor and similarity encoder while keeping the remaining modules fixed. 
The adaptation stage used a learning rate of \(1\times10^{-3}\) and a batch size of 4. 
To evaluate adaptation efficiency, we varied the number of patient-specific synthetic samples and compared the results with patient-specific training from random initialization.

\begin{figure}[h!]
    \centering
        \includegraphics[width=\linewidth]{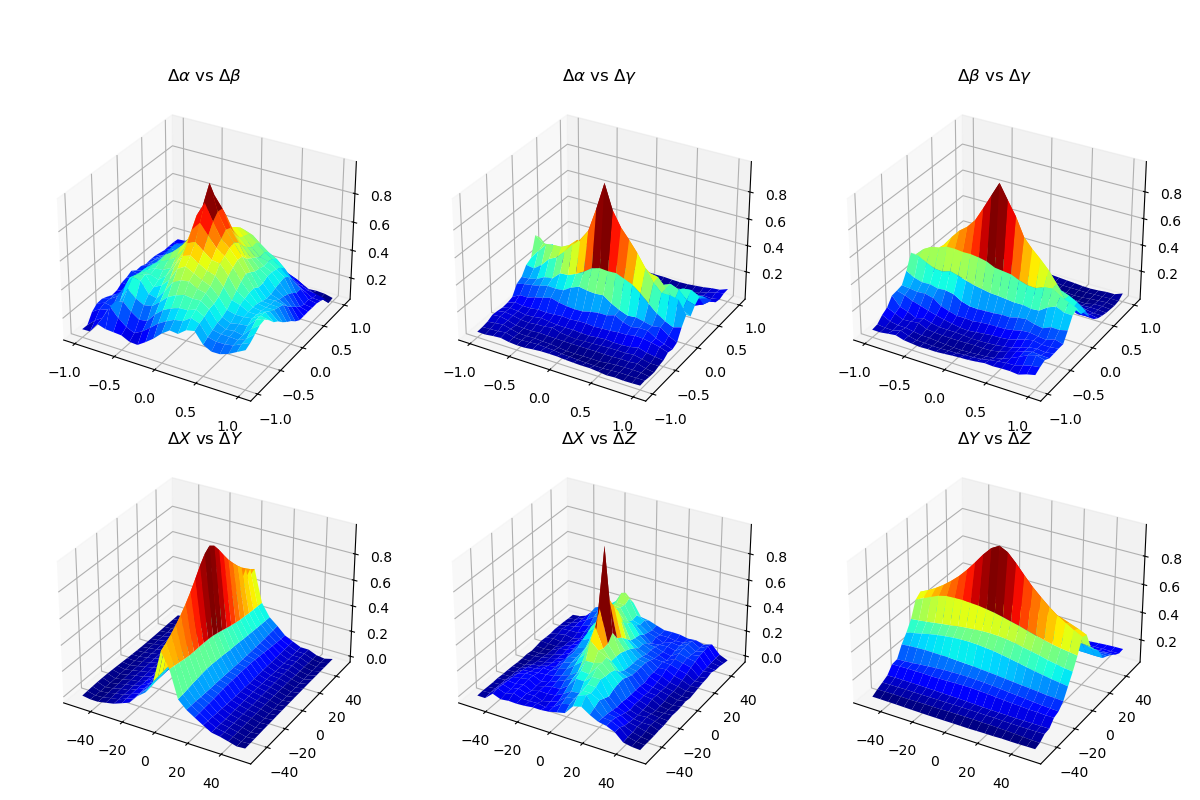}
\caption{Visualization of the proposed spherical deep similarity landscape.
For clearer visualization, the deep similarity values are first normalized to the range [0,1], and then transformed by computing 1-$\epsilon$, effectively inverting the scale to enhance contrast in the display.}
\label{fig::similarity}
\end{figure}

At inference time, the adapted pose regressor first predicted an initial pose, which was then refined using differentiable Levenberg--Marquardt optimization. 
The LM optimizer was run for 150 iterations with the learning rate initialized to \(5\times10^{-3}\). 
A step learning-rate decay with a factor of 0.9 was applied every 25 iterations. 
The termination criterion was defined such that the standard deviation of the learned similarity over the last ten iterations was less than \(1\times10^{-2}\). 
The final pose after LM refinement was used for quantitative evaluation. 
All input DRRs and X-ray images were resized to \(256 \times 256\) and normalized to \([0,1]\).
The learned deep similarity is visualized in Fig.~\ref{fig::similarity}
\subsection{Efficient Patient-Specific Adaptation}
\label{sec:efficient_adaptation}
\begin{table*}[h!]
\centering
\caption{
Ablation study of fine-tuned modules during patient-specific adaptation.
All settings use the same number of patient-specific synthetic DRRs.
}
\label{tab:module_ablation}
\resizebox{\textwidth}{!}{
\begin{tabular}{lcccccc}
\toprule
Dataset
& Regressor
& \makecell{Similarity\\network}
& \makecell{E-CNN \\ module}
& SMSR $\uparrow$
& \makecell{Median\\mTRE (mm) $\downarrow$}
& \makecell{Training\\time (min)} \\
\midrule

\multirow{4}{*}{DeepFluoro}
& \checkmark & \xmark & \xmark
& 68.5\% & 0.86 & 7.4 \\
& \xmark & \checkmark & \xmark
& 76.8\% & 0.68 & 18.7 \\
& \checkmark & \checkmark & \xmark
& 83.9\% & 0.57 & 25.6 \\
& \checkmark & \checkmark & \checkmark
& \textbf{84.7\%} & \textbf{0.55} & 28.3 \\

\midrule

\multirow{4}{*}{Ljubljana}
& \checkmark & \xmark & \xmark
& 71.0\% & 0.78 & 7.8 \\
& \xmark & \checkmark & \xmark
& 79.5\% & 0.63 & 19.2 \\
& \checkmark & \checkmark & \xmark
& 84.0\% & 0.54 & 25.1 \\
& \checkmark & \checkmark & \checkmark
& \textbf{85.0\%} & \textbf{0.52} & 28.9 \\

\bottomrule
\end{tabular}
}
\end{table*}
The comparison results between the proposed method and other patient-specific baseline methods are presented in Table~\ref{tab::patient-specific}, the baselines follow the same setting as we introduced in~\cite{chen2026intraoperative}.
Table~\ref{tab:main_adaptation} compares different patient-specific adaptation strategies on DeepFluoro and Ljubljana, while Fig.~\ref{fig::training} further shows the corresponding training dynamics. 
The patient-specific model trained from scratch serves as the upper-bound baseline, whereas the patient-agnostic-only model evaluates direct transfer without target-patient adaptation. 
Direct deployment of the patient-agnostic model leads to a clear performance drop, achieving only 19.6\% SMSR on DeepFluoro and 23.4\% on Ljubljana. 
This indicates that although patient-agnostic pretraining learns reusable representations, patient-specific adaptation remains necessary to account for target anatomy and dataset-specific appearance differences.
Patient-specific fine-tuning rapidly recovers registration accuracy. 
On DeepFluoro, SMSR increases from 19.6\% without adaptation to 74.3\% with 50\% fine-tuning, and reaches 84.7\% with full fine-tuning, close to the from-scratch result of 86.1\%. 
At the same time, training time is reduced from 20.70 hours to 0.47 hours, corresponding to an approximately 44-fold speed-up. 
On Ljubljana, 10\% fine-tuning already improves SMSR from 23.4\% to 70.0\%, and full fine-tuning matches the from-scratch SMSR of 85.0\% while reducing training time from 20.90 hours to 0.48 hours. 

As shown in Fig.~\ref{fig::training}, training from scratch improves gradually over more than 20 hours, whereas patient-specific fine-tuning reaches near from-scratch performance within a short adaptation period. 
These results demonstrate that patient-agnostic synthetic pretraining provides an effective initialization for patient-specific 2D/3D registration, converting a costly from-scratch training process into a lightweight adaptation problem while preserving comparable registration accuracy.
\subsection{Adaptation Module Analysis}
\label{sec:module_analysis}
Motivated by recent advances in parameter-efficient fine-tuning of vision foundation models~\cite{kong2026pro,lin2024beyond,chen2024ma}, we investigate multiple patient-specific adaptation strategies by the set of modules updated during fine-tuning. This allows us to analyze the trade-off between registration accuracy and adaptation cost.
Table~\ref{tab:module_ablation} analyzes which modules should be updated during patient-specific adaptation. 
Fine-tuning only the pose regressor already improves performance compared with direct patient-agnostic transfer, reaching 68.5\% SMSR on DeepFluoro and 71.0\% on Ljubljana. 
This indicates that adapting the initialization network to the target anatomy is beneficial. 
However, fine-tuning only the similarity network achieves higher accuracy, with SMSR increasing to 76.8\% on DeepFluoro and 79.5\% on Ljubljana, suggesting that the learned similarity landscape also requires patient-specific adaptation.
Updating both the pose regressor and the similarity network provides the best trade-off between accuracy and efficiency. 
On DeepFluoro, this setting achieves 83.9\% SMSR and a median mTRE of 0.57 mm, which is close to full fine-tuning with 84.7\% SMSR and 0.55 mm median mTRE. 
Similarly, on Ljubljana, fine-tuning the regressor and similarity network reaches 84.0\% SMSR and 0.54 mm median mTRE, while full fine-tuning achieves 85.0\% SMSR and 0.52 mm median mTRE. 
The additional adaptation of the E-CNN module brings only marginal improvement but increases training time.

These results suggest that most of the patient-specific adaptation benefit comes from updating the pose initialization and similarity estimation components. 
Therefore, fine-tuning the regressor and similarity network while keeping the volume module fixed offers an efficient adaptation strategy with near full fine-tuning performance.
% \subsection{Sample Efficiency of Patient-Specific Adaptation}

\subsection{Effect of Segmentation-Free Domain Randomization}
Table~\ref{tab:dr_ablation} evaluates the effect of different domain randomization strategies during patient-agnostic pretraining. 
We compare five settings: no domain randomization (No DR), the randomization strategy used in our conference version (Previous DR), intensity and contrast randomization only (\(\mathcal{G}_{I}\)), intensity/contrast plus projection-physics randomization (\(\mathcal{G}_{I}+\mathcal{G}_{P}\)), and the proposed full segmentation-free domain randomization strategy. 
The proposed strategy further includes the mask, occlusion, and field-of-view pool \(\mathcal{G}_{M}\) during the synthetic pretraining process, in addition to \(\mathcal{G}_{I}\) and \(\mathcal{G}_{P}\).
\begin{table}[h!]
\centering
\caption{
Ablation of domain randomization strategies during patient-agnostic pretraining. Median and 95\% denote mTRE statistics.
}
\label{tab:dr_ablation}
\begin{tabular}{lccc}
\toprule
Setting 
& SMSR $\uparrow$ 
& Median $\downarrow$ 
& 95\% $\downarrow$ \\
\midrule
\multicolumn{4}{l}{\textit{DeepFluoro}} \\
No DR        & 76.2\% & 0.70 & 2.18 \\
Previous DR~\cite{chen2026intraoperative}  & 82.7\% & 0.59 & 1.70 \\
\(\mathcal{G}_{I}\) & 79.6\% & 0.64 & 1.91\\
\(\mathcal{G}_{I} + \mathcal{G}_{P}\) & 83.2\% & 0.58 & 1.65 \\
\(\mathcal{G}_{I} + \mathcal{G}_{P}+ \mathcal{G}_{M}\)  & \textbf{84.7\%} & \textbf{0.55} & \textbf{1.58} \\
\midrule
\multicolumn{4}{l}{\textit{Ljubljana}} \\
No DR        & 77.0\% & 0.70 & 2.20 \\
Previous DR~\cite{chen2026intraoperative}   & 83.0\% & 0.56 & 1.55 \\
\(\mathcal{G}_{I}\)  & 80.0\% & 0.62 & 1.84 \\
\(\mathcal{G}_{I} + \mathcal{G}_{P}\) & 83.8\% & 0.55 & 1.47 \\
\(\mathcal{G}_{I} + \mathcal{G}_{P}+ \mathcal{G}_{M}\) & \textbf{85.0\%} & \textbf{0.52} & \textbf{1.33} \\
\bottomrule
\end{tabular}
\end{table}
Examples of the proposed segmentation-free domain randomization are provided in Fig.~\ref{fig::examples}
Without domain randomization, the pretrained model achieves 76.2\% SMSR on DeepFluoro and 77.0\% on Ljubljana, indicating that synthetic pretraining alone is insufficient to fully bridge the DRR-to-X-ray appearance gap. 
Using only intensity and contrast randomization improves performance to 79.6\% and 80.0\% SMSR, respectively, showing that photometric perturbations help reduce sensitivity to fixed DRR intensity distributions. 
Adding projection-physics randomization further improves the results to 83.2\% on DeepFluoro and 83.8\% on Ljubljana, suggesting that acquisition-related variations such as detector shifts, zoom changes, and bias fields are important for improving synthetic-to-real transfer.

Compared with the previous domain randomization strategy, the proposed full randomization achieves consistently better performance. 
On DeepFluoro, it improves SMSR from 82.7\% to 84.7\% and reduces the 95th percentile mTRE from 1.70 mm to 1.58 mm. 
On Ljubljana, it improves SMSR from 83.0\% to 85.0\% and reduces the 95th percentile mTRE from 1.55 mm to 1.33 mm. 
These results indicate that the mask/occlusion/FOV perturbations in \(\mathcal{G}_{M}\) provide complementary robustness beyond intensity and projection-physics randomization for pretraining. 
Overall, the proposed segmentation-free domain randomization better diversifies the synthetic pretraining distribution and improves patient-specific adaptation without requiring anatomical segmentation masks or landmark annotations.

% \section{Discussion}\label{sec12}

% Discussions should be brief and focused. In some disciplines use of Discussion or `Conclusion' is interchangeable. It is not mandatory to use both. Some journals prefer a section `Results and Discussion' followed by a section `Conclusion'. Please refer to Journal-level guidance for any specific requirements. 

\section{Conclusion}
\label{sec:conclusion}

In this work, we proposed an efficient patient-specific 2D/3D registration framework based on patient-agnostic synthetic pretraining and patient-specific adaptation. 
The model first learns transferable pose-sensitive representations from synthetic DRRs generated across multiple CT volumes, and is then adapted to a target patient using a limited number of patient-specific synthetic projections. 
We further introduced a segmentation-free domain randomization strategy to improve synthetic-to-real robustness without requiring anatomical labels or segmentation masks.
Experiments on DeepFluoro and Ljubljana showed that direct patient-agnostic transfer is insufficient, while patient-specific fine-tuning rapidly recovers registration accuracy. 
Compared with training from scratch, the proposed pretraining--adaptation strategy achieved comparable accuracy while reducing patient-specific training time by more than 40 times. 
Ablation studies further demonstrated the benefits of fine-tuning the pose regressor and similarity network, as well as the effectiveness of the proposed domain randomization.
Future work will validate the method on larger multi-center clinical datasets and investigate faster adaptation strategies for real-time intraoperative deployment. 
Overall, our results suggest that patient-agnostic synthetic pretraining can transform patient-specific 2D/3D registration from a costly from-scratch training problem into an efficient adaptation problem.

\backmatter

\bmhead{Data availability}
We used two publicly available datasets for 2D/3D registration: DeepFluoro
(\url{https://github.com/rg2/DeepFluoroLabeling-IPCAI2020}) and Ljubljana
(\url{https://lit.fe.uni-lj.si/en/research/resources/3D-2D-GS-CA}).

\bmhead{Author contribution}
M. C.: Conceptualization, Methodology, Software, Validation, Formal analysis,
Investigation, Data curation, Visualization, Writing -- original draft, Writing -- review
and editing. 
Y. K.: Supervision, Resources,
Project administration, Writing -- review and editing.
\section*{Declarations}
\noindent{\textbf{Conflict of interest}} The authors declare no competing interests.

\bibliography{reference}% common bib file

\end{document}